\documentclass[twocolumn]{aastex62}
\usepackage[utf8]{inputenc}
\usepackage{longtable}
\usepackage{graphicx}
\usepackage{amsmath,amssymb, fixmath}
\usepackage{color, colortbl}
\usepackage{units}
\usepackage{epstopdf}
\usepackage{hyperref}
\usepackage{multirow}
\usepackage{url}
\usepackage{subfigure}
\usepackage{rotating}
\usepackage[nobiblatex]{xurl}
\usepackage{enumitem}\setlist[description]{font=\textendash\enskip\scshape\bfseries}
\usepackage{lineno}
\definecolor{amber}{rgb}{1.0, 0.75, 0.0}

\begin{document}

\title{Bayesian model selection for GRB 211211A through multi-wavelength analyses}

\author[0000-0002-1275-530X]{Nina Kunert}\affiliation{Institute of Physics and Astronomy, Theoretical Astrophysics, University Potsdam, Haus 28, Karl-Liebknecht-Str. 24/25, 14476, Potsdam, Germany}
\author[0000-0002-7686-3334]{Sarah Antier}\affiliation{Artemis, Observatoire de la Côte d’Azur, Université Côte d’Azur, Boulevard de l'Observatoire, 06304 Nice, France}
\author[0000-0002-5196-2029]{Vsevolod Nedora}\affiliation{Max Planck Institute for Gravitational Physics (Albert Einstein Institute), Am Mühlenberg 1, Potsdam 14476, Germany}
\author[0000-0002-8255-5127]{Mattia Bulla}\affiliation{Department of Physics and Earth Science, University of Ferrara, via Saragat 1, I-44122 Ferrara, Italy}\affiliation{INFN, Sezione di Ferrara, via Saragat 1, I-44122 Ferrara, Italy}\affiliation{INAF, Osservatorio Astronomico d’Abruzzo, via Mentore Maggini snc, 64100 Teramo, Italy}
\author[0000-0001-7041-3239]{Peter T. H. Pang}\affiliation{Nikhef, Science Park 105, 1098 XG Amsterdam, The Netherlands}\affiliation{Institute for Gravitational and Subatomic Physics (GRASP), Utrecht University, Princetonplein 1, 3584 CC Utrecht, The Netherlands}
\author[0000-0003-3768-7515]{Shreya Anand}\affiliation{Cahill Center for Astrophysics, California Institute of Technology, Pasadena CA 91125, USA}
\author[0000-0002-8262-2924]{Michael W. Coughlin}\affiliation{School of Physics and Astronomy, University of Minnesota, Minneapolis, Minnesota 55455, USA}
\author[0000-0003-2656-6355]{Ingo Tews}\affiliation{Theoretical Division, Los Alamos National Laboratory, Los Alamos, NM 87545, USA}
\author[0000-0003-3340-4784]{Jennifer Barnes}\affiliation{Kavli Institute for Theoretical Physics, Kohn Hall, University of California, Santa Barbara, CA 93106, USA}
\author[0009-0009-2434-432X]{Thomas Hussenot-Desenonges}\affiliation{IJCLab, Univ Paris-Saclay, CNRS/IN2P3, Orsay, France}
\author[0000-0002-7718-7884]{Brian Healy}\affiliation{School of Physics and Astronomy, University of Minnesota, Minneapolis, Minnesota 55455, USA}
\author[0009-0003-6181-4526]{Theophile Jegou du Laz}\affiliation{Division of Physics, Mathematics, and Astronomy, California Institute of Technology, Pasadena, CA 91125, USA}
\author{Meili Pilloix}\affiliation{Artemis, Observatoire de la Côte d’Azur, Université Côte d’Azur, Boulevard de l'Observatoire, 06304 Nice, France}
\author[0000-0002-9108-5059]{Weizmann Kiendrebeogo}\affiliation{Artemis, Observatoire de la Côte d’Azur, Université Côte d’Azur, Boulevard de l'Observatoire, 06304 Nice, France}\affiliation{Laboratoire de Physique et de Chimie de l’Environnement, Université Joseph KI-ZERBO, 03 BP 7021, Ouagadougou, Burkina Faso}
\author[0000-0003-2374-307X]{Tim Dietrich}\affiliation{Institute of Physics and Astronomy, Theoretical Astrophysics, University Potsdam, Haus 28, Karl-Liebknecht-Str. 24/25, 14476, Potsdam, Germany}\affiliation{Max Planck Institute for Gravitational Physics (Albert Einstein Institute), Am Mühlenberg 1, Potsdam 14476, Germany}

\date{\today}

\begin{abstract}

Although GRB 211211A is one of the closest gamma-ray bursts (GRBs), its classification is challenging because of its partially inconclusive electromagnetic signatures.
In this paper, we investigate four different astrophysical scenarios as possible progenitors for GRB~211211A: 
a binary neutron-star merger, 
a black-hole--neutron-star merger, 
a core-collapse supernova, and an r-process enriched core collapse of a rapidly rotating massive star (a collapsar). 
We perform a large set of Bayesian multi-wavelength analyses based on different models describing these scenarios and priors to investigate which astrophysical scenarios and processes might be related to GRB~211211A. 
Our analysis supports previous studies in which the presence of an additional component, likely related to $r$-process nucleosynthesis, is required to explain the observed light curves of GRB~211211A, as it can not solely be explained as a GRB afterglow. 
Fixing the distance to about $350~\rm Mpc$, namely the distance of the possible host galaxy SDSS J140910.47+275320.8, we find a statistical preference for a binary neutron-star merger scenario.
\end{abstract}

\section{Introduction}

The joint detection of gravitational waves (GWs) and electromagnetic (EM) signatures originating from the merger of binary neutron stars (BNSs)
on August 17th 2017~(\citealt{LVCGW170817,LSCMM2017ApJ}) was a breakthrough in multimessenger astronomy. 
In addition to the GW signal GW170817, an associated kilonova, AT2017gfo, and a gamma-ray burst (GRB), GRB~170817A, were observed~\citep{LSCMM2017ApJ}. 
This multimessenger detection allowed for an independent way of measuring the expansion rate of the Universe~\cite{LIGOScientific:2017adf}, 
placed new constraints on the properties of supranuclear-dense matter~\citep{Bauswein:2017vtn,Ruiz:2017due,Radice:2017lry,Most:2018hfd,Coughlin:2018fis,Capano:2019eae,Dietrich2020,Huth:2021bsp}, and proved that at least some short GRBs are connected to compact binary mergers~\citep{2017ApJ...848L..13A}. 
However, it was also reported that short GRBs could originate from collapsars \citep{2021NatAs...5..917A}, indicating that the classification of astrophysical scenarios associated with GRBs is more complex (\cite{Zhang:2021agu,Rossi:2021bau}).
Additional signatures associated with GRBs and their afterglows, such as kilonovae, help significantly in the identification of the origin of the progenitors. 
The kilonova AT2017gfo was certainly an exemplary case for such an EM signal, and spectral features connected to the creation of new elements~\citep{WaHa19,Domoto:2022cqp} in the outflowing material have possibly been observed. 
In addition to AT2017gfo, there are a large number of kilonova candidates that could be connected to other GRB observations, e.g., GRB~060614,  GRB~130603B,  GRB~150101B, GRB~150424A, GRB~160821B, GRB~060505 e.g., \cite{Tanvir:2013pia,Berger:2013wna,YaJi2015,JinLi2015,FoMa2016,TrRy2018, Lamb:2019lao,KaKoLau2017,JinWang2018, Troja:2019ccb, Jin:2019uqr}; cf. e.g.~\cite{Ascenzi:2018mbh} for a review about some of these kilonova candidates. 
The most recent example that has to be added to the list is the kilonova candidate associated with GRB~211211A.

This GRB signal was discovered on the 11th December 2021 at 13:09:59 (UT) with the Burst Alert Telescope (BAT) of the Swift Observatory, with its optical and near-infrared counterpart observed by, for example, \cite{Rastinejad:2022zbg} and \cite{Troja:2022yya}. 

This GRB signal is characterized by a complex emission phase lasting approximately 50~s and shows several overlapping pulses lasting for about $\sim$~12~s (\cite{Rastinejad:2022zbg}).
Given this duration, GRB~211211A would be classified as a long GRB typically arising from the core-collapse of a massive star (e.g., \citealt{Stanek_2003,2016SSRv..202...33L}) and not from compact binary mergers. 
Hence, for a scenario such as GRB~211211A, one would not necessarily expect to observe an associated kilonova. 
Based on intensive follow-up observations \citep{Rastinejad:2022zbg, Troja:2022yya}, it seems plausible that SDSS~J140910.47+275320.8 was the host galaxy of GRB~211211A, at 98.6 percent confidence~\citep{Rastinejad:2022zbg}.

Numerous groups, e.g., \citep{Rastinejad:2022zbg, Troja:2022yya, Yang:2022qmy, Mei:2022ncd}, and also other groups explained these observations by invoking a kilonova in associated with GRB~211211A. 
This was suggested for various reasons: 
(i) the profile of the prompt emission showed an initially complex structure followed by an extended softer emission, 
(ii) a predominant signature of a supernova was lacking for up to 17 days post-discovery, 
(iii) the color evolution of the optical counterpart had similar properties as AT2017gfo, and 
(iv) the offset of the GRB location concerning the center of the host galaxy was larger than for typical long GRBs.
 
The origin of GRB~211211A is hotly debated: \cite{Yang:2022qmy}, for example, suggested that it has similar properties as GRB~060614, another event associated with a kilonova candidate. They conclude that the significant excess in the near-infrared and optical afterglow at late observations points more towards a neutron star-white dwarf merger that leaves behind a rapidly spinning magnetar as a central engine providing additional heating to the ejecta.

\cite{2022arXiv220511112S} mentioned a possible gamma-ray precursor before the main emission, which was caused by the resonant shattering of one star’s crust prior to the merger. 
In contrast, \cite{2022arXiv220505031G} argued for the presence of a strong magnetic field from the precursor surrounding the central engine of the GRB.
This would result in the prolongation of the accretion process and, thus, could explain the duration of the hard spiky emission detected for GRB~211211A. 
Similarly, \cite{2022arXiv220502186X} supposes that a magnetar participated in the merger and caused a quasi-periodic precursor.
\cite{2022arXiv220505008G} analysed the spectra of the prompt emission of GRB~211211A by using synchrotron spectrum models and concluded that the rapid evolution of synchrotron emission was the main driver its extended emission. While the kilonova observed for GRB~211211A argues for a BNS merger, a neutron-star--black-hole (NSBH) scenario cannot be fully ruled out. 
Finally, \cite{Barnes:2023} investigated the possibility that collapsars could explain the origin of GRB~211211A and found that the afterglow-subtracted emission of GRB~211211A is in best agreement for collapsar models with high kinetic energies.

Following the discussion in the literature, we will use our nuclear physics and multimessenger astrophysics (\texttt{NMMA}) framework~\citep{Pang:2022rzc}\footnote{\url{https://github.com/nuclear-multimessenger-astronomy}} to explore various astrophysical scenarios for the origin of GRB~211211A.
We will consider the possibility of two merger scenarios, a BNS merger and an NSBH merger, and in addition two supernova scenarios, a core-collapse supernova, and an $r$-process enriched collapsar. At this point, we emphasize that while multiple scenarios (e.g. different supernova types) could possibly explain the origin of GRB~211211A, we restrict our study to the four scenarios mentioned above, and to particular models representing such scenarios. Hence, our study will only provide estimates for this narrow parameter space of possible scenarios.
For our model selection study, the \texttt{NMMA} framework allows us to simultaneously fit the observed data across the full electromagnetic range with multiple models; for example, we can simultaneously employ GRB afterglow and kilonova models without the need to split the observational data in chunks and processing them separately.

\section{Observational data }\label{sec:data}

In order to perform our model selection, we collect a set of multi-wavelength data observed for GRB~211211A. However, we do not use any data from the prompt emission phase of the GRB in our analysis because our framework is, at the current stage, not able to handle such highly energetic emission.

With regard to the X-ray data, we use the available information from the \textit{Swift} X-ray Telescope. In particular, we use the 0.3~-~10~keV~flux light curve observed at late times ($t = 10^4$~s after BAT trigger time) and convert it to 1-keV~flux~densities following \cite{2008ApJ...689.1161G}. For our optical study in UV, we use results from \textit{Swift}-UVOT in table~2 provided by \cite{Rastinejad:2022zbg} in the bands $v$, $b$, $u$, $uvw1$, $uvm2$, $uvw2$, and $white$. To supplement these UVOT data, we incorporate measurements from ``supplement Table 1" provided by \citet{Troja:2022yya}. Whenever measurements overlap within a 30-min window in the same band between \citet{Troja:2022yya} and \cite{Rastinejad:2022zbg}, we remove duplicates, as such measurements represent variations in analysis binning during the early epoch between the two publications.\footnote{Moreover, we ensure in this way that the statistical analysis will not be biased by including redundant data, which would give more weight to specific bands during the likelihood estimation.} For the remaining optical data, we exclusively utilize measurements directly analysed by \cite{Rastinejad:2022zbg} (cf. table~1 in the appendix, references 1).
We follow a similar approach for the published data from \citet{Troja:2022yya}. 
(cf. supplement table 1). 
Furthermore, we augment our data set with measurements exclusively published in the General Coordinates Network \href{https://gcn.gsfc.nasa.gov/other/211211A.gcn3}{(GCNs)} \citep{2021GCN.31233....1P,2021GCN.31232....1M} for instance. 
In this manner, we try to obtain an almost complete set encompassing available optical data for this particular GRB, including the latest publicly accessible measurements when possible. The data were all corrected from the foreground Galactic extinction A$_V$ = 0.048 mag~\citep{SchaflyFinkbeiner2011}.

Furthermore, we use the 6~GHz radio detection of GRB~211211A observed 6.27 d after the initial trigger with a 5$\sigma$ upper limit flux density of 16~$\mu$Jy \citep{Rastinejad:2022zbg}. With regard to available GeV data, as reported in \cite{Zhang:2022fzj} and \cite{Mei:2022ncd}, we do not include this data because our employed GRB model does not provide mechanisms to explain their origin.

\section{Methods}\label{sec:methods}

\subsection{Bayesian Inference}

Our analysis is based on the nuclear physics and multimessenger astronomy framework \texttt{NMMA}~\citep{Pang:2022rzc} that allows us to perform joint Bayesian inference runs of multimessenger events containing GWs, kilonovae, supernovae, and GRB afterglow signatures. 
For this paper, we extended the code infrastructure to include the description of $r$-process enriched collapsars following the model of~\cite{Barnes_2022}.

We use the EM data of GRB~211211A to investigate which model or which combination of models describe the observational data best.
According to Bayes' theorem, we compute posterior probability distributions, $p(\textbf{\textit{$\theta$}} | d, M)$, for model source parameters \textbf{\textit{$\theta$}} under the hypothesis or model $M$ with data $d$ as
\begin{equation}\label{Eq:Bayes_theorem}
\begin{aligned}
    p(\textbf{\textit{$\theta$}} | d, M) = \frac{p(d|\textbf{\textit{$\theta$}},M) p(\textbf{\textit{$\theta$}} | M)}{p(d|M)}\to
    \mathcal{P}(\textbf{\textit{$\theta$}} ) = \frac{\mathcal{L}(\textbf{\textit{$\theta$}}) \pi(\textbf{\textit{$\theta$}})}{\mathcal{Z}(d)}\,,
\end{aligned}
\end{equation}
where $\mathcal{P}(\textbf{\textit{$\theta$}})$, $\mathcal{L}(\textbf{\textit{$\theta$}})$, $\pi(\textbf{\textit{$\theta$}})$, and $\mathcal{Z}(d)$ are the posterior, likelihood, prior, and evidence, respectively. In order to investigate the plausibility of competing models, we evaluate the odds ratio $\mathcal{O}^{1}_{2}$ for two models $\mathbf{M}_1$ and $\mathbf{M}_2$ which is given by
\begin{equation}
\mathcal{O}^{1}_{2}  = \frac{p(d | M_1)}{p(d | M_2)}\frac{p(M_1)}{p(M_2)} \equiv \mathcal{B}^{1}_{2}\Pi^{1}_{2},
\end{equation}
where $\mathcal{B}^{1}_{2}$ and $\Pi^{1}_{2}$ are the Bayes factor and the prior odds, respectively.
Under the assumption that the different astrophysical scenarios considered here are equally likely to explain GRB~211211A, we impose unity prior odds, namely $\Pi^1_2$ = 1, for all comparisons of models describing these scenarios.
Therefore, we simply compute the Bayes factor $\mathcal{B}^1_2$. 
In our study, we report the natural logarithm of the Bayes factor,
\begin{equation}\label{Eq:Bayes_Factor}
    \ln \mathcal{B}^1_{\rm{ref}} = \ln\left(\frac{p(d|M_1)}{p(d|M_{\rm{ref}})}\right)\,,
\end{equation}%
relative to our best fitting model as a reference (ref.), which we will denote as $\ln \mathcal{B}_{\rm{ref}}$ hereafter.
Following \cite{Jeffreys1961} and \cite{Kass1995}, we interpret $\ln \mathcal{B}^1_{\rm{ref}} $ as the evidence favoring our reference model as:
  \begin{tabular}{m{4.2cm} m{3.7cm}}
    $ \ln [\mathcal{B}^1_{\rm{ref}}] < -4.61$& decisive evidence, \\
    $-4.61 \leq \ln [\mathcal{B}^1_{\rm{ref}}] \leq -2.30$  & strong evidence, \\
    $-2.30 \leq \ln [\mathcal{B}^1_{\rm{ref}}] \leq -1.10$  & substantial evidence, \\
    $-1.10 \leq \ln [\mathcal{B}^1_{\rm{ref}}] \leq 0$  & no strong evidence. \\
  \end{tabular}
However, we point out that these classifications should only be considered as estimates and that the Bayes factor is generally a continuous quantity.
In addition to the Bayes factor, we also provide information about the ratio of the maximum likelihood, or the difference of the maximum log-likelihood point estimates $\ln [\mathcal{L}^1_2 (\hat{\theta})]$ supporting our analysis in Sec.~\ref{subsec:model_comparison}.
We will denote this as $\ln [\mathcal{L}_{\rm ref} (\hat{\theta})]$ when we compare the maximum log-likelihood against our reference model. 

\subsection{Employed models}\label{empl_models}

As described in the Introduction, we investigate four different scenarios in our study from which GRB~211211A could have emerged. In particular, we consider two merger scenarios: a BNS merger and an NSBH merger, and two supernova cases, namely a phenomenological long GRB supernova template and an r-process enriched collapsar scenario. As a word of caution, we emphasize that all these scenarios can only be considered when employing characteristic models describing such a scenario. 

Strictly speaking, our results will only favor or disfavor particular models employed for the analysis, but we will not be able to rule out an entire astrophysical scenario; for example, disfavoring the employed SN98bw and $r$CCSNe will not be sufficient to rule out the supernova origin completely.

\textbf{BNS scenario}: For this case, we use the kilonova models of \cite{Dietrich2020} (hereafter `BNS-KN-Bulla') and of \cite{Kasen:2017sxr} (hereafter `BNS-KN-Kasen'). BNS-KN-Bulla is based on the time-dependent 3D Monte Carlo radiation transfer code \texttt{possis} (\cite{Bulla:2019muo, Bulla:2022mwo}), which computes light curves, spectra, and luminosities for kilonovae depending on the viewing-angle $\theta_{\rm{Obs}}$. 
The ejected material is classified through the dynamical ejecta mass, $M_{\rm{ej}}^{\rm{dyn}}$, and the disc-wind ejecta mass, $M_{\rm{ej}}^{\rm{wind}}$.
The tidal dynamical ejecta component is assumed to be distributed within a half-opening angle $\Phi$. In the same way, BNS-KN-Kasen uses the multi-dimensional Monte Carlo code \texttt{sedona}, which solves the multiwavelength radiation transport equation in a relativistically expanding medium (\cite{Kasen:2006ce, Roth:2014wda}). 
In this paper, we use the 1D model provided by \cite{Kasen:2017sxr}, which assumes spherical symmetry and uniform composition for our analysis. The model, `BNS-KN-Kasen', depends on the ejecta mass, $M_{\rm{ej}}$, a characteristic expansion velocity, $v_{\rm{ej}}$, and the mass fraction of lanthanides, $X_{\rm{lan}}$, which affects the opacity.

\textbf{NSBH scenario}: For this case, we also use a \texttt{possis} model grid of kilonova spectra tailored to NSBH mergers which was used in the study of \cite{Anand:2020eyg} (hereafter `NSBH-KN-Bulla'). This model depends on the same model parameters as BNS-KN-Bulla but excludes the dependence on the half opening angle of the dynamical ejecta, fixed to $\Phi=30^\circ$.

\textbf{Supernova}: In order to assess the possibility of a typical core-collapse supernova (CCSN) associated with a long GRB, we use the \texttt{nugent-hyper} model from \texttt{sncosmo}~\citep{Levan:2004sn} with the absolute magnitude, $S_{\rm{max}}$, as the main free parameter. This model is a template constructed from observations of the supernova SN1998bw associated with the long GRB~980425 and is hereafter abbreviated as `SN98bw'.

\textbf{$r$-process enriched Collapsar}: Rapidly rotating massive star core collapses \citep{RevModPhys.29.547, Qian:1996xt} are another possible astrophysical site for $r$-process nucleosynthesis. 
As massive stars undergo a core collapse, material is disrupted and forms an accretion disk which can become neutron-rich through weak interactions \citep{Beloborodov:2002af}, and can launch winds that power emission of $r$-process-enriched core-collapse SNe ($r$CCSNe).
We use the semi-analytic model for $r$CCSNe of \cite{Barnes_2022} (hereafter denoted as 'SNCol'). 
The model depends on five free parameters: the total ejecta mass, $M_{\rm{ej}}$; a characteristic ejecta velocity, $v_{\rm{ej}}$; the $^{56}$Ni mass, $M_{\rm{Ni}}$; the $r$-process material mass, $M_{\rm{rp}}$; and the mixing coordinate, $\Psi_{\rm{mix}}$. 
The ejecta are assumed to be spherically symmetric, with $r$-process elements of mass $m_{\rm{rp}}$ concentrated in an inner core whose total mass is $\Psi_{\rm{mix}} m_{\rm{ej}}$, with $\Psi_{\rm{mix}} \leq 1$. An $r$-process-free envelope surrounds the core, and 56-Ni is distributed uniformly throughout the core and the envelope. The velocity $v_{\rm{ej}}$ is defined such that the total kinetic energy of the ejecta $E_{\rm{kin}}$ is equal to $\frac{1}{2} M_{\rm{ej}} v_{\rm{ej}}^2$.\footnote{\cite{Barnes:2023} also compared $r$CCSNe with observational data from GRB~211211A. However, not within a Bayesian approach as employed here and with an updated version of their model originally described in \cite{Barnes_2022}.}

\textbf{GRB afterglow}: For modeling the GRB afterglow light curves, we employ the semi-analytic model of \cite{vanEerten:2010zh} and \cite{Ryan:2019fhz}, available in the public \texttt{afterglowpy} library (denoted as `GRB-M'). 
The model computes GRB afterglow emission and takes the following free parameters as input:
the isotropic kinetic energy, $E_{\rm K,iso}$; the viewing angle, $\theta_{\rm{Obs}}$; the half-opening angle of the jet core, $\theta_c$; the outer truncation angle of the jet, $\theta_w$; the interstellar medium density, $n$; the electron energy distribution index, $p$; and the fractions of the shock energy that go into electrons, $\epsilon_e$ and magnetic fields, $\epsilon_B$.
The model allows for several angular structures of the GRB jet. 
For our simulations, we assume a Gaussian or a top-hat jet structure (hereafter, `Gauss' and  `top')\footnote{In addition, we tested a power law jet structure for which we found consistent results.}. 
It is important to note that, while we try to be agnostic concerning the origin of GRB~211211A, the GRB-M model that we employ has some limitations. 

Specifically, it does not include the emission from the reverse shock, which might be important at early times.
Additionally, it does not include the wind-like interstellar medium, which is expected in the case of a collapsar.

\begin{figure*}
    \centering
    \includegraphics[width=0.95\textwidth]{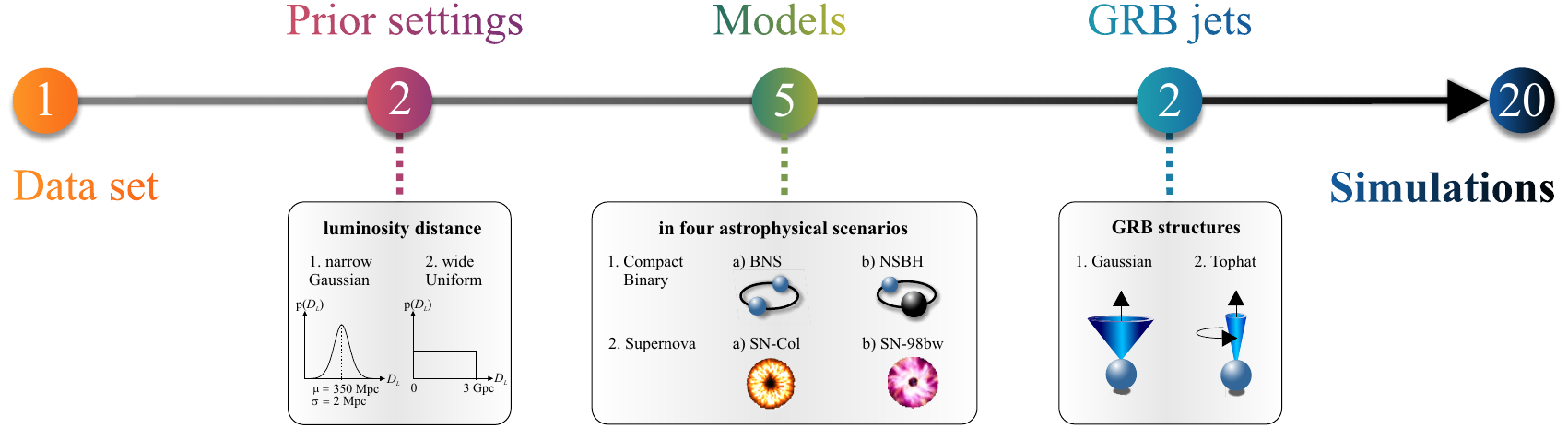}
    \caption{Schematic illustration of our comprehensive Bayesian inference campaign performed to analyze GRB~211211A. We use one observational data set as described in Sec.~\ref{sec:data}; two prior settings in which we mainly vary the luminosity distance prior while prior settings for other model parameters remained fixed and are reported in Table~\ref{tab:Prior_bounds_Dfix}; five models (including two different BNS kilonova models) or model combinations for four different astrophysical scenarios; and two GRB jet types (Gaussian and top-hat), resulting in 20 Bayesian inferences.
    }
    \label{fig:infographic}
\end{figure*}

In Fig.~\ref{fig:infographic}, we summarize our approach to analysing GRB~211211A based on the data set described in Sec.~\ref{sec:data}. 
We employ two different priors for the luminosity distance, namely a narrow Gaussian luminosity distance prior centered around 350~Mpc as reported by \cite{Rastinejad:2022zbg} and a uniform prior on the luminosity distance ranging between $0$ and 3~Gpc.
This allows us to investigate the potential influence of the distance on the GRB classification. 
Furthermore, we employ five models or model combinations to describe the different astrophysical scenarios.
For the choice of a Gaussian luminosity distance prior, we report the prior settings for all parameters of the employed models in Table \ref{tab:Prior_bounds_Dfix}. 
Moreover, we use two different GRB jet types, resulting in 20 Bayesian inference simulations.

\section{Multi-wavelength Analyses}\label{sec:MMA}

In the following three subsections, \ref{subsec:model_comparison}-\ref{subsec:merger_props}, we discuss our results for a narrow Gaussian prior on the luminosity distance in order to compare with previous studies. 
In subsection \ref{subsec:results_inf_settings}, we will investigate the influence of the distance prior choice and employ a wide uniform prior on the luminosity distance. 

\begin{table*}[t]
  \centering
  \begin{tabular}{|m{3.1cm}| m{3.7cm}| m{1.5cm} |m{1.3cm}| m{2.0cm} | m{1.6cm} |} 
    \toprule
    Name  & Astrophysical & GRB Jet & Model & Bayes factor & Likelihood \\ 
     & Processes  & Structure & dimension & $ \ln[\mathcal{B}_{\rm{ref}}^1]$ & $\ln [\mathcal{L}^1_{\rm{ref}} (\hat{\theta})]$ \\ \hline 

    BNS-GRB-M$^{\rm Kasen}_{\rm top}$ & Kilonova + GRB  & Tophat & 11 &  ref. & ref. \\
    BNS-GRB-M$^{\rm Kasen}_{\rm Gauss}$ & Kilonova + GRB  & Gaussian & 12 & -1.21	 $\pm$ 0.12  & 0.04 \\
    BNS-GRB-M$^{\rm Bulla}_{\rm top}$  & Kilonova + GRB  & Tophat & 11 & -4.51	 $\pm$ 0.12  &  -3.24	 \\ 
    BNS-GRB-M$^{\rm Bulla}_{\rm Gauss}$ & Kilonova + GRB  & Gaussian & 12 & -6.26 $\pm$ 0.12 & -3.31 \\ \hline
    NSBH-GRB-M$_{\rm top}$ & Kilonova  + GRB & Tophat &  11 & -8.41 $\pm$ 0.12  &  -9.30 \\
    NSBH-GRB-M$_{\rm Gauss}$  & Kilonova + GRB & Gaussian & 12 & -10.56 $\pm$ 0.12 & -8.99 \\ \hline 
    SNCol-GRB-M$_{\rm top}$& $r$CCSNe + GRB& Tophat & 14 & -15.24 $\pm$ 0.13  & -8.41 \\ 
    SNCol-GRB-M$_{\rm Gauss}$ & $r$CCSNe + GRB & Gaussian & 15 &  -16.97 $\pm$ 0.13 & -8.34 \\ \hline 
    SN98bw-GRB-M$_{\rm top}$   & CCSNe + GRB & Tophat  & 8 &  -12.66 $\pm$ 0.12  & -14.69 \\
    SN98bw-GRB-M$_{\rm Gauss}$ & CCSNe + GRB & Gaussian & 9 &  -12.59	$\pm$ 0.12 & -13.01\\ \hline 
    GRB-M$_{\rm top}$  & GRB  & Tophat & 8 & -12.47	 $\pm$ 0.12 & -13.01  \\ 
    GRB-M$_{\rm Gauss}$ & GRB & Gaussian & 9 & -12.65 $\pm$ 0.12 & -14.67 \\
  \hline    
  \end{tabular}
  \caption{Results for the logarithmic Bayes factors, $\ln[\mathcal{B}^1_{\rm{ref}}]$, and maximum logarithmic likelihood ratios, $\ln [\mathcal{L}^1_{\rm{ref}}(\hat{\theta})]$, relative to the best-fitting, joint inference using BNS-GRB-M$^{\rm Kasen}_{\rm top}$ (ref.). The four investigated scenarios of possible astrophysical origins (BNS, NSBH, SNCol, and SN98bw) are each assessed assuming a Gaussian or a top-hat jet structure. 
  As reference, we list results for a stand-alone GRB-M investigation for both jet structures.}
  \label{tab:odds_ratio}
\end{table*}

\subsection{Model Comparison}\label{subsec:model_comparison}

As indicated in the Introduction, one of the main differences between previous studies and our work is that most previous works fitted first the X-ray and radio data with a GRB afterglow model, and then used the afterglow-subtracted optical and NIR photometry for fitting a kilonova model. 
In contrast, but similar to \cite{Yang:2022qmy}, we perform a joint analysis of the GRB afterglow and a possible additional contribution such as a kilonova signature or emission from a $r$CCSN or CCSN. Moreover, in order to consider systematic uncertainties arising from different assumptions made in each model, we employ a 1~mag uncertainty in our simulations.

\begin{figure*}[h!]
    \centering
    \includegraphics[width=\textwidth]{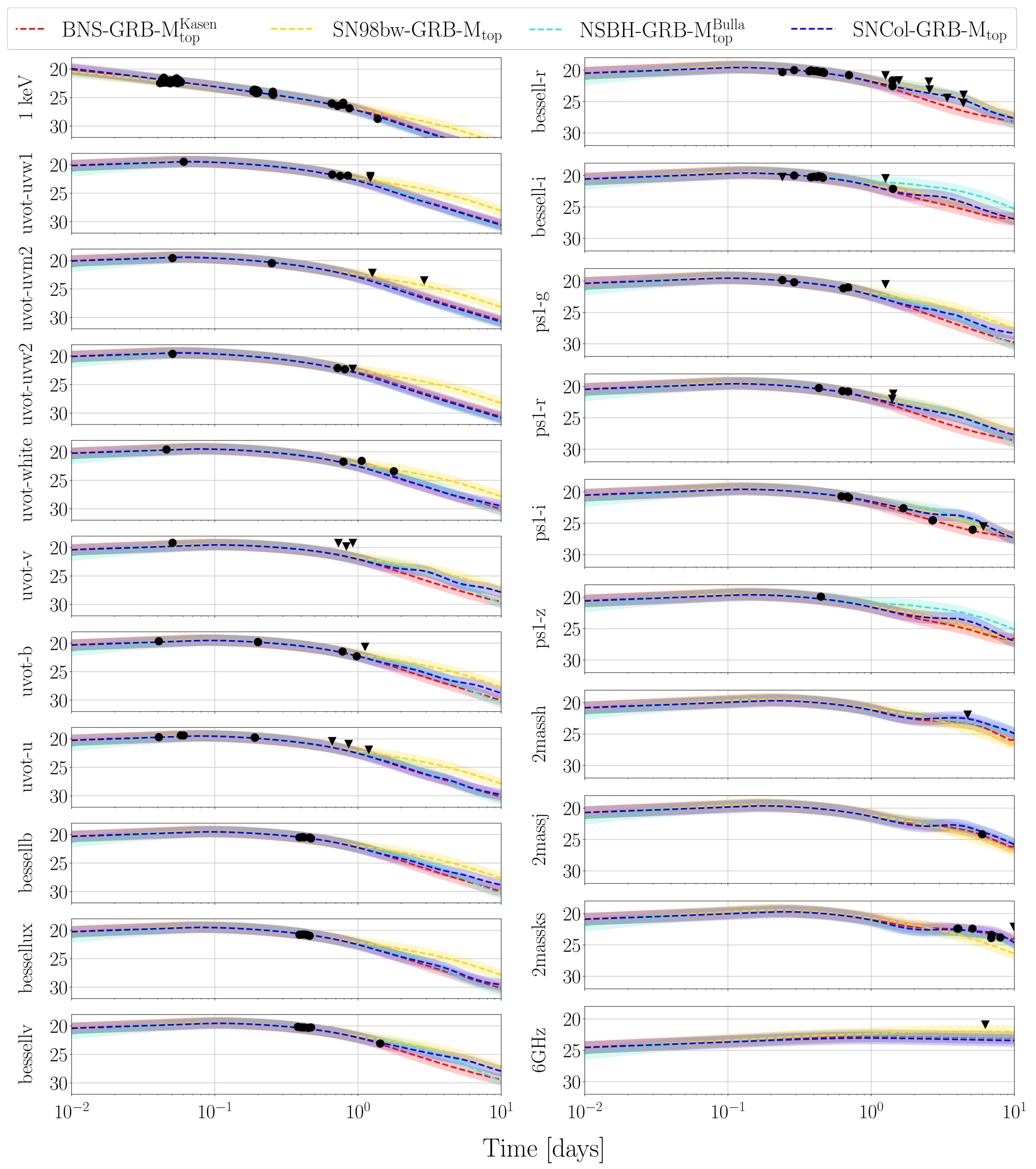}
    \caption{Best-fitting light curve from joint Bayesian inferences listed in Table~\ref{tab:odds_ratio} for possible scenarios: BNS-GRB-M$^{\rm Kasen}_{\rm top}$ (red), NSBH-GRB-M$_{\rm top}$ (cyan), SNCol-GRB-M$_{\rm top}$ (orange), and SN98bw-GRB-M$_{\rm top}$ (blue). The observational data of GRB~211211A in X-ray-1keV, radio-6GHz, UV, optical, and NIR band as discussed in Sec.~\ref{sec:data} are shown as black dots, whereas black triangles refer to upper detection limits.  Note that we are employing the naming convention of the \href{https://sncosmo.readthedocs.io/en/stable/bandpass-list.html}{\texttt{sncosmo}} library for our work.}
    \label{fig:comb_bands}
\end{figure*}

In Table~\ref{tab:odds_ratio}, we summarize our main findings for the investigated astrophysical scenarios. 
We found that the BNS-GRB-M$^{\rm Kasen}_{\rm top}$ model describes the observational data best, and hence we pick it as our reference model. 
Consequently, the Bayes factors and likelihood ratios in Table~\ref{tab:odds_ratio} are reported relative to this best-fitting inference run. With reference to Table \ref{tab:odds_ratio}, we show the maximum log-likelihood light-curve fits in Fig.~\ref{fig:comb_bands} for each assessed scenario, which we will refer to as 'best-fitting light curves' hereafter.

Comparing only the two different BNS kilonova models, we find differences in the log-Bayes factors of about 4, disfavouring the BNS-GRB-M$^{\rm Bulla}_{\rm Gauss/top}$ model compared to our reference model. 
Different GRB afterglow models lead only to a change of about 1 in the log-Bayes factor. Similarly, the employed GRB afterglow model has only a very small imprint of the maximum log-likelihood values, while different kilonova models lead to a change of order 3. 
These differences can be seen in Fig.~\ref{fig:lcs_compare_BNS_vs_GRBalone} especially in the bands \textit{bessel-v}, \textit{ps1-i}, \textit{2massj}, and \textit{2massks}. Although the maximum likelihood light curve for the BNS-GRB-M$^{\rm Kasen}_{\rm Gauss}$ simulation is slightly favored as compared to our reference scenario, the difference is of the order of the statistical uncertainties.

It is worth pointing out that statistical uncertainties, as stated in the table, are noticeably smaller than model differences; that is, our results are dominated by systematic uncertainties in the underlying light-curve models. 

Considering the differences between the NSBH and BNS scenarios, we find strong evidence that GRB~211211A was connected to a BNS rather than an NSBH system. This is reflected both in Bayes factors as well as maximum log-likelihood values as shown in Table~\ref{tab:odds_ratio}. Comparing the respective best-fitting light curves in Fig.~\ref{fig:comb_bands}, we see that NSBH-GRB-M$^{\rm Bulla}_{\rm top}$ fits the NIR-band data worse compared with BNS-GRB-M$^{\rm Kasen}_{\rm top}$.

With regard to the relative Bayes factors for the collapsar scenario, we find that there is decisive evidence that a BNS scenario is preferred over a collapsar origin for GRB~211211A when employing the light-curve models outlined in the previous section.
However, it is important to note that the collapsar model depends on more parameters. 
Because of this, Occam's razor penalizes the model.

As indicated by \cite{Rastinejad:2022zbg}, and confirmed by our study, we find that a Ni-powered SN event or an SN98bw-GRB-M scenario is noticeably less favored compared to a BNS merger. This is depicted in Fig.~\ref{fig:comb_bands} in which SN98bw-GRB-M$_{\rm top}$ fails to fit late-time NIR data, resulting in a larger, negative log-likelihood ratio. Moreover, the upper SN limits reported for the $r$- and $i$-band in \citet{Troja:2022yya} rule out other supernova models.

Finally, our study confirms that the BNS-GRB-M$^{\rm Kasen}_{\rm top}$ scenario provides decisive evidence when compared with GRB-M$_{\rm top}$ simulations, even though the latter sampled over fewer parameters in the respective parameter estimation runs. 
Considering the impact of the choice of a Gaussian vs.\ top-hat jet structure on our Bayes factor results, we find a slight preference for the top-hat jet structure for all assessed scenarios, except for SN98bw-GRB-M$_{\rm top}$.

\subsection{Presence of an additional component}\label{subsec:add_messenger}

\begin{figure}[t]
    \centering
    \includegraphics[width=0.47\textwidth]{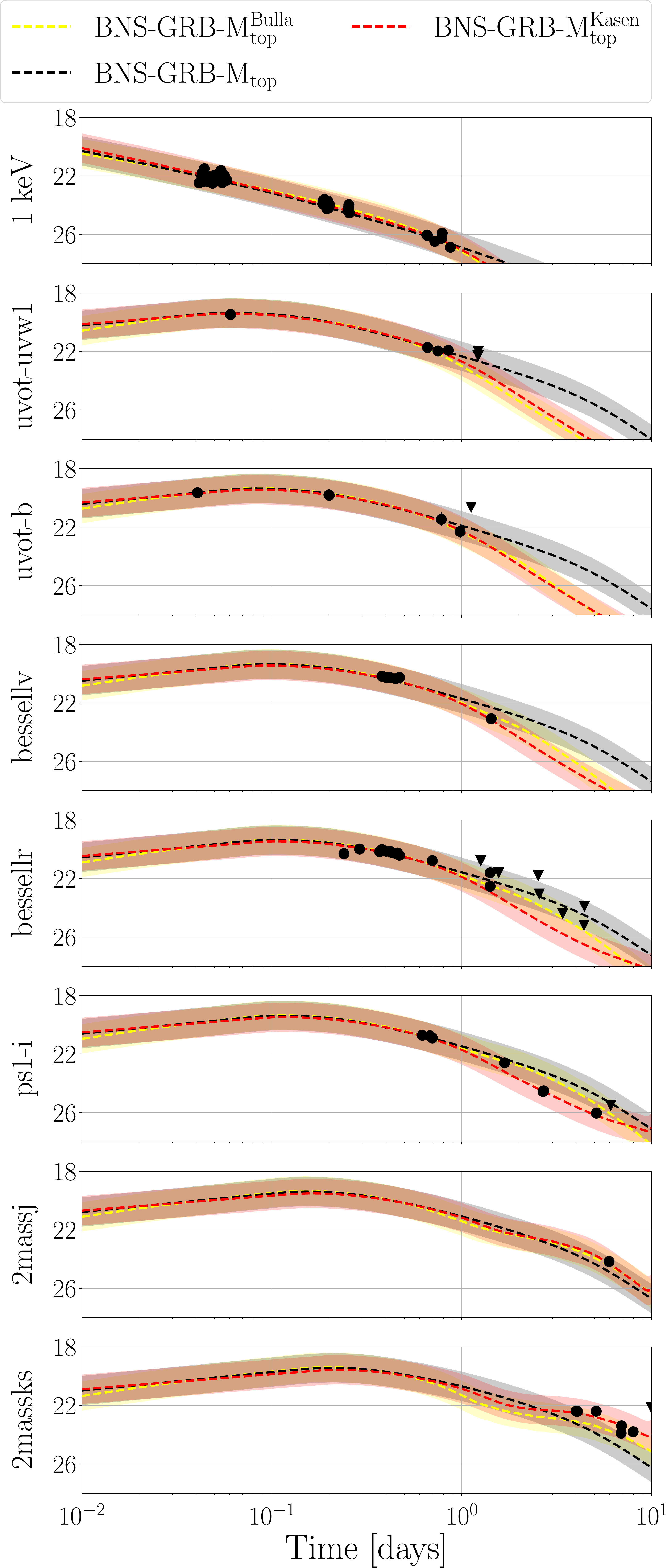}
    \caption{Best-fitting light curves from joint Bayesian inferences of BNS-GRB-M$^{\rm Bulla}_{\rm top}$ (yellow) and BNS-GRB-M$^{\rm Kasen}_{\rm top}$ (red) compared to a stand-alone GRB-M$_{\rm top}$ inference (black) for X-ray, UV, optical and NIR bands on a logarithmic time scale in days since trigger time.}
    \label{fig:lcs_compare_BNS_vs_GRBalone}
\end{figure}

Given the overall narrative that GRB~211211A was a GRB connected to a kilonova, we study the ability of the GRB-M with top-hat jet structure to describe the observational data and compare this with two BNS merger scenarios. For this purpose, we show the best-fitting light curves for BNS-GRB-M$^{\rm Bulla}_{\rm top}$, BNS-GRB-M$^{\rm Kasen}_{\rm top}$, and GRB$_{\rm top}$ in Fig.~\ref{fig:lcs_compare_BNS_vs_GRBalone} for a selection of the most informative bands.

We find that the GRB-M achieves a good representation of the data in the X-ray and UV bands (not all are shown in Fig.~3). However, the optical bands such as \textit{uvot-b} and \textit{bessel-v} already show that an additional component is required to describe the dimmer observational data observed roughly 1 d after trigger time. This becomes even more pronounced in NIR bands, especially in the \textit{ps1-i} and \textit{2massks} bands, where our reference model achieves the best representation.
Overall, the joint model inferences of BNS-GRB-M$^{\rm Bulla}_{\rm top}$ and BNS-GRB-M$^{\rm Kasen}_{\rm top}$ achieve a better representation in the mentioned bands and the observational data points lie within the estimated 1 magnitude uncertainty (shaded band) of the best-fitting light curves.
Hence, our analysis suggests that an additional source of energy generation is required to generate bright light curves in optical and NIR bands and to fit the observed data. 

\subsection{Source properties of the potential compact binary mergers}\label{subsec:merger_props}
 
For the scenario that GRB~211211A was connected to a compact binary merger, which is favored by our analysis, we now determine the source properties of the potential progenitor system. 
For this purpose, we use the inferred GRB afterglow and kilonova properties for both BNS-KN-Kasen and BNS-KN-Bulla and connect information about the ejecta and debris disc to the BNS properties following~\cite{Dietrich2020}; cf.~\cite{Henkel:2022naw} for a recent discussion about uncertainties in the employed numerical relativity-informed phenomenological relations.

\begin{figure}[t]
    \centering
    \includegraphics[width=0.5\textwidth]{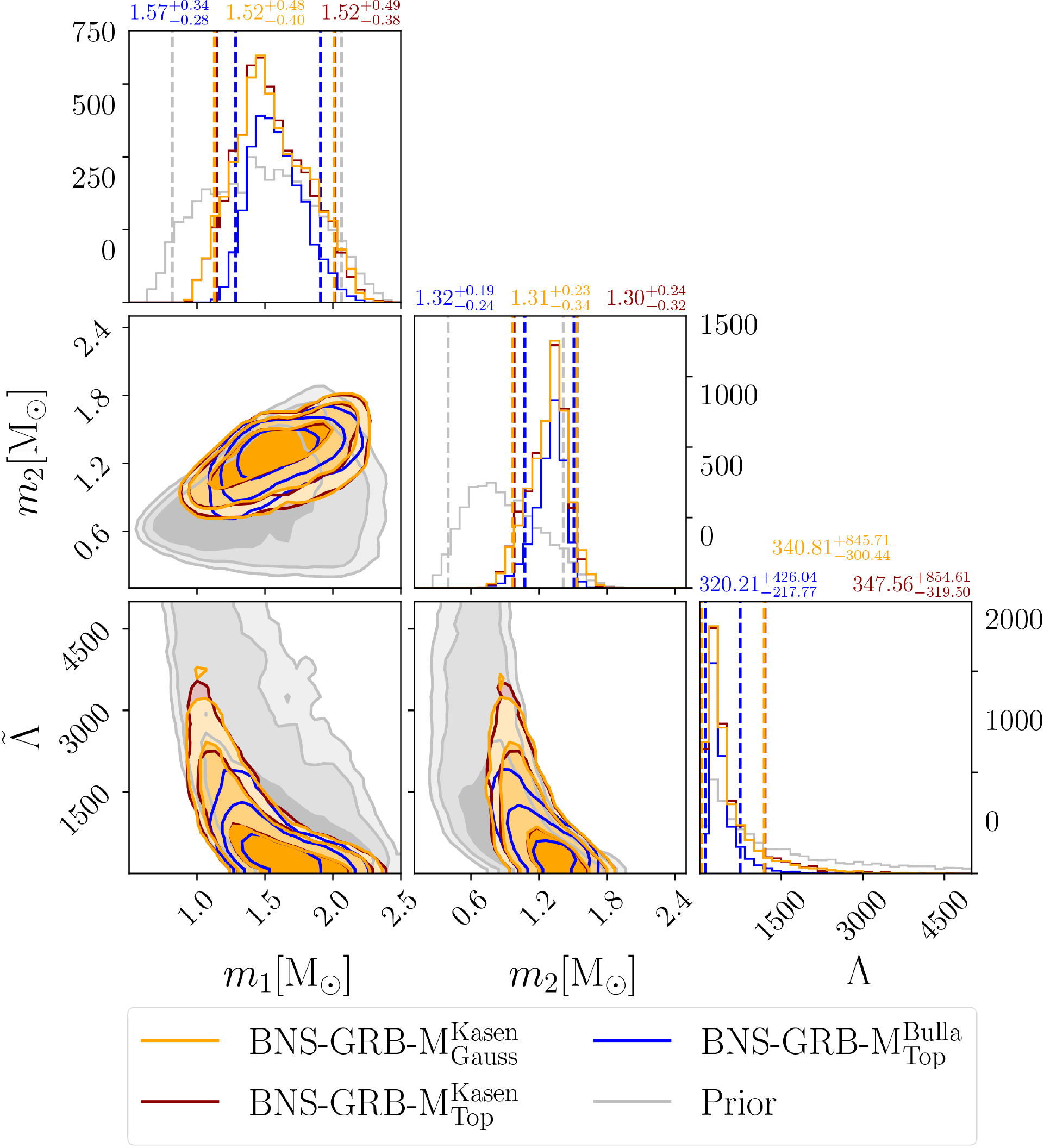}
    \caption{Component masses $m_{1,2}$ and the dimensionless tidal deformability $\rm{\tilde{\Lambda}}$ based on our inference results of BNS-GRB-M$^{\rm{Kasen}}_{\rm{Gauss}}$ (orange), BNS-GRB-M$^{\rm{Kasen}}_{\rm{Top}}$ (red) and BNS-GRB-M$^{\rm{Bulla}}_{\rm{Top}}$ (blue). Different shadings mark the 68 per cent, 95 per cent, and 99 per cent confidence intervals. 
    For the 1D posterior probability distributions, we give the 90 per cent confidence interval (dashed lines) and report median values above each panel. 
    Grey shaded areas give the prior probability regions.}\label{fig:corner_GWEMresampled-masses-lamT}
\end{figure}

In Fig.~\ref{fig:corner_GWEMresampled-masses-lamT}, we show our inference results for a possible BNS source using BNS-GRB-M$^{\rm Kasen}_{\rm top}$, BNS-GRB-M$^{\rm Kasen}_{\rm Gauss}$, and BNS-GRB-M$^{\rm Bulla}_{\rm top}$ and contrast these with the prior probability regions for each parameter, in order to show how constraining the observational data are.  
Comparing inference results for BNS-GRB-M$^{\rm Kasen}_{\rm Top}$ and BNS-GRB-M$^{\rm Kasen}_{\rm Gauss}$, we find that estimated source masses and tidal deformabilities are very similar.
For the BNS-GRB-M$^{\rm Kasen}_{\rm Top}$ simulation, we find that a BNS merger with a primary mass of 1.52$^{+0.49}_{-0.38} M_{\odot}$ and a secondary mass of 1.30$^{+0.24}_{-0.32} M_{\odot}$ was the likely progenitor. The associated dimensionless tidal deformability of the system lies within $\tilde{\Lambda} = 348^{+855}_{-320}$. 
With regard to a similar analysis for BNS-GRB-M$^{\rm Bulla}_{\rm Top}$, we find a primary mass of 1.57$^{+0.34}_{-0.28} M_{\odot}$ and a secondary mass of 1.32$^{+0.19}_{-0.24} M_{\odot}$. The corresponding tidal deformability is $320^{+426}_{-218}$. Comparing estimated masses for BNS-GRB-M$^{\rm Kasen}_{\rm Top}$ and BNS-GRB-M$^{\rm Bulla}_{\rm Top}$, we find overall good agreement within the stated uncertainties. 

Overall, our estimated masses are consistent with \cite{Rastinejad:2022zbg}, who concluded that GRB~211211A originated from a 1.4~$M_{\odot}$+1
.3~$M_{\odot}$ BNS merger. 
We expect that the remaining small differences are caused by the different analyses of the observed GRB~211211A data and by the fact that \cite{Rastinejad:2022zbg} assumed the inclination angle, under which the binary was observed, to be zero. 
Moreover, \cite{Rastinejad:2022zbg} assumed a fixed equation of state (EOS) from the EOS set of \cite{Dietrich2020} using additional information from~\cite{Nicholl:2021rcr}. 
In contrast, we leave the inclination angle as a free parameter in our analysis and use the updated EOS set of~\cite{Huth:2021bsp}.
This set incorporates information from theoretical nuclear physics computations and from astrophysical observations of neutron stars such as \cite{Dietrich2020}, but also heavy-ion collision experimental data. 
With regard to investigated binary merger scenarios, we find that the inferred inclination angle is around $\theta_{\rm{Obs}} \approx 0.02^{+0.02}_{-0.02}$rad, while a larger inclination angle of $\theta_{\rm{Obs}} \approx 0.04^{+0.03}_{-0.02}$rad is estimated for the considered supernova scenario (see Table \ref{tab:Prior_bounds_Dfix}). 

\cite{Rastinejad:2022zbg} deduced a total $r$-process ejecta mass of $M_{\rm{ej}} = 0.047^{+0.026}_{-0.011} M_{\odot}$, of which $0.02~M_{\odot}$ correspond to lanthanide-rich ejecta, $0.01~M_{\odot}$ to intermediate-opacity ejecta, and $0.01~M_{\odot}$ to lanthanide-free material. In addition, \cite{Yang:2022qmy} reported a total ejecta mass of $M_{\rm{ej}} = 0.037^{+0.008}_{-0.004} \rm{M}_{\odot}$.
With our reference inference result from BNS-GRB-M$^{\rm Kasen}_{\rm top}$, we find a total ejecta mass of $M^{\rm{BNS}}_{\rm ej, Kasen}~=~0.016^{+0.013}_{-0.009} \rm{M}_{\odot}$, which is smaller than the result estimated by \cite{Rastinejad:2022zbg}. Concerning our analysis based on BNS-GRB-M$^{\rm Bulla}_{\rm top}$, we found a total ejecta mass of $M^{\rm{BNS}}_{\rm ej, Bulla} = 0.022^{+ 0.021}_{-0.013} \rm{M}_{\odot}$, of which $0.012 \rm{M}_{\odot}$ can be attributed to lanthanide-rich ejecta, $0.006 \rm{M}_{\odot}$ to intermediate opacity mass, and $0.001 \rm{M}_{\odot}$ to lanthanide-free material. We note that during our analysis of GRB~211211A, we have also performed simulations not including the \textit{Swift}-UVOT data. In such a case, we find larger total ejecta masses of $M^{\rm{BNS}}_{\rm ej, Kasen}~=~0.021^{+0.017}_{-0.013} \rm{M}_{\odot}$ and $M^{\rm{BNS}}_{\rm ej, Bulla} = 0.031^{+ 0.033}_{-0.018} \rm{M}_{\odot}$ comparable to \cite{Rastinejad:2022zbg} and \cite{Yang:2022qmy}. This finding sheds some light on the impact of different data sets being used to analyse astronomical sources such as GRB~211211A in a Bayesian context.

For completeness, we have performed a similar investigation for our NSBH-GRB-M$_{\rm top}$ and NSBH-GRB-M$_{\rm Gauss}$ models to infer the corresponding NSBH properties by making use of the relations provided in ~\cite{Foucart_2018} and \cite{Kruger:2020gig}.
Although the observational data do not provide a strong constraint on the NSBH source properties, our NSBH-GRB-M$_{\rm top}$ analysis suggests that an NSBH merger with a BH mass of 3.11$^{+5.53}_{-2.23} M_{\odot}$ and an NS mass of 1.40$^{+0.74}_{-0.81} M_{\odot}$ could have been the progenitor of GRB~211211A, with a total ejecta mass of $M^{\rm{NSBH}}_{\rm ej}~=~0.006^{+0.006}_{-0.004}~M_{\odot}$.
Likewise, the BH spin is weakly constrained to $\chi_1 = 0.00^{+0.59}_{-0.60}$ for the NSBH-GRB-M$_{\rm{top}}$ inference.  
Our inferred NS masses are in agreement with previous GW population analyses \citep{LIGOScientific:2018mvr, LIGOScientific:2020ibl, LIGOScientific:2021djp} and with the maximum non-spinning NS mass of $2.7^{+0.5}_{-0.4} M_{\odot}$ estimated at 90 per cent credibility by \cite{Ye:2022qoe}. 
Within the estimated uncertainties, the inferred BH mass is close to the NSBH mass gap for which the lightest BH masses were estimated to be $\sim 5 M_{\odot}$ \citep{ozel_2010, Farr_2011}. 

\subsection{Influence of the prior choice}\label{subsec:results_inf_settings}

Finally, we discuss the influence of a different luminosity distance prior on our results. 
The distance of GRB~211211A was relatively precisely estimated based on the redshift of the potential host galaxy, $z~=~0.0763~\pm~0.0002$ \citep{Rastinejad:2022zbg}. However, we are generally interested in the influence of a wide uniform luminosity distance prior on our results.
For this reason, we widen the prior range and allow a distance between 0 and 3~$\rm Gpc$. 

Following the procedure in Sec.~\ref{subsec:model_comparison}, we have computed the logarithmic Bayes factors and found that BNS-GRB-M$^{\rm Kasen}_{\rm top}$ remains to be the best-fitting model. Moreover, the differences in logarithmic Bayes factors between BNS-KN-Bulla and BNS-KN-Kasen remain the same. 
The Bayes factors are slightly smaller for all assessed scenarios when comparing to the results in Tab.~\ref{tab:odds_ratio}. Interestingly, the SN98bw-GRB-M$_{\rm{Gauss/top}}$ and the GRB-M$_{\rm{Gauss/top}}$ are least favored, while the NSBH and collapsar scenario are more favored. Overall, our main conclusions remain valid also for the wider distance prior.

\begin{figure}[t]
    \centering
    \includegraphics[width=0.44\textwidth]{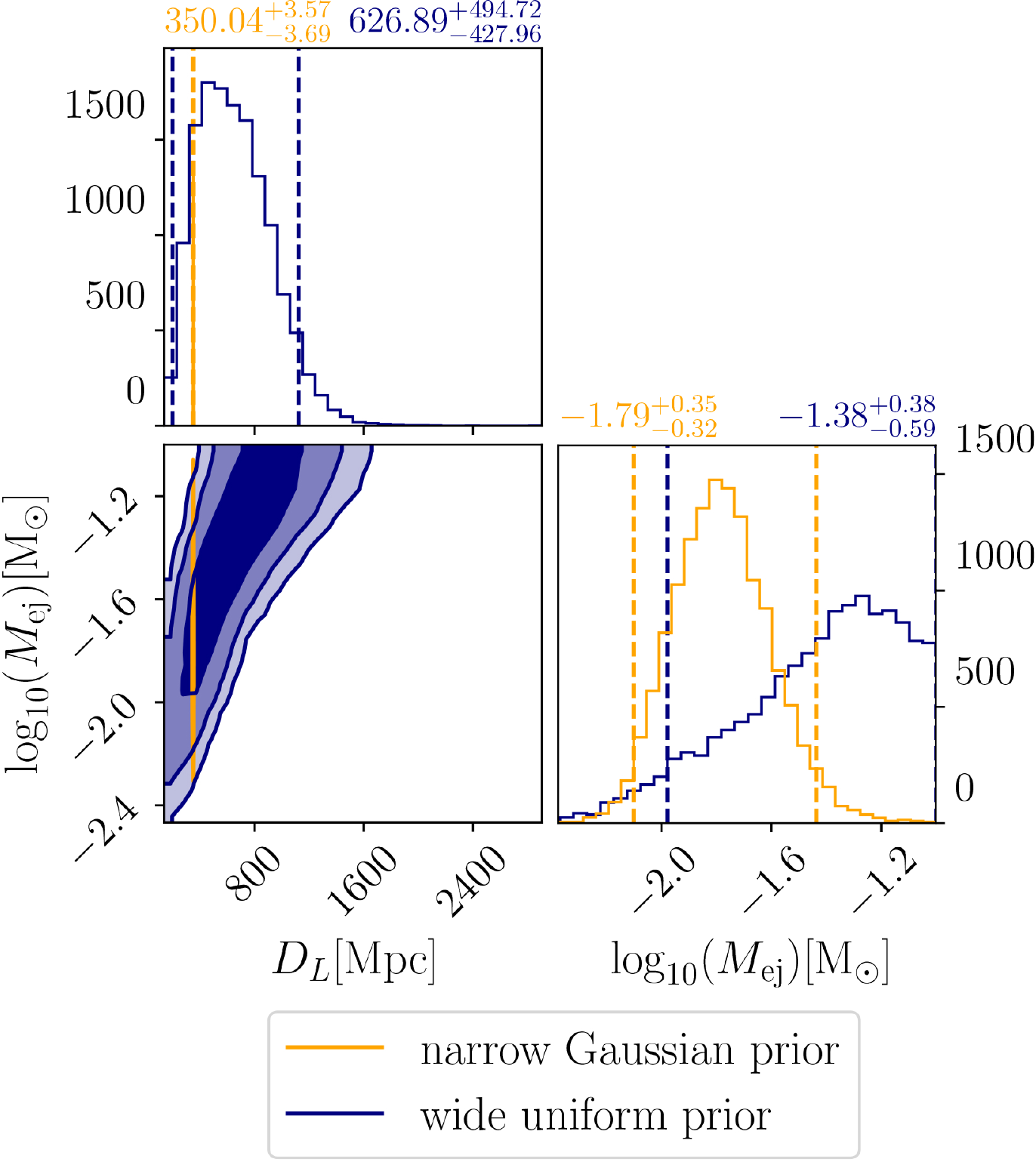}
    \caption{Corner plot for BNS-GRB-M$^{\rm{Kasen}}_{\rm{top}}$ with a narrow Gaussian luminosity distance prior centered around 350~Mpc (orange) and a wide uniform luminosity distance prior ranging up to 3~Gpc (blue). 
    The inferred model parameters are shown at 68 per cent, 95 per cent, and 99 per cnet confidence (shadings from light to dark). 
    For the 1D posterior probability distributions,
    we report the median values and show the 90 per cent confidence intervals as dashed lines.}\label{fig:Corner_BNS_narrowwideDL}
\end{figure}

We investigated the posterior probability distributions obtained for a wide uniform distance prior and compared these with the ones obtained for a narrow Gaussian distance prior setting. 
In Fig.~\ref{fig:Corner_BNS_narrowwideDL}, we show an example for the obtained luminosity distance and the total ejecta mass distributions using BNS-GRB-M$^{\rm{Kasen}}_{\rm{top}}$. 
As can be seen, the wide distance prior leads to a noticeably weaker constraint on the distance and the total ejecta mass. 
The latter is caused by a degeneracy between the luminosity distance and the ejecta mass. 
Generally, larger ejecta masses could compensate for larger distances and vice versa, which explains the shape of the 2D correlation plot of Fig.~\ref{fig:Corner_BNS_narrowwideDL}.
Similarly (not shown in the figure), also the SNCol model predicts higher ejecta masses for larger distances. 
With respect to the SN-GRB and the GRB inferences, the GRB isotropic energy, $\log_{\rm{10}} (E_{\rm{K, iso}})$, tends to increase for larger distances, which is expected as brighter signals can be detected at further distances.

\section{Conclusion}\label{sec:Conclusion}

In this paper, we have performed multiple multiwavelength analyses for GRB~211211A assuming four different scenarios, namely a BNS merger, an NSBH merger, an $r$CCSN, as well as a CCSN. 
On the basis of joint multiwavelength Bayesian inferences combining respective kilonova or SN models with a GRB afterglow model, we investigated which which gave the strongest statistical evidence to explain the data detailed in Sec. \ref{sec:data}. While emphasizing again that our study only considers a small proportion of possible astrophysical scenarios, and thus our results need to be considered with caution, we summarize our main conclusions below.

\begin{enumerate}[label=(\roman*)]
\item On the basis of the four assessed scenarios and the employed models, we find statistical evidence for a BNS merger scenario; cf.\ Table~\ref{tab:odds_ratio}. However, we can not fully rule out other scenarios. 
\item Our study confirms that GRB~211211A can not solely be explained as a GRB afterglow and that an additional emission process (likely related to $r$-process nucleosynthesis) is required for a good description of the observational data, mostly in optical and NIR bands (cf.~Fig.~\ref{fig:lcs_compare_BNS_vs_GRBalone}). This emphasizes that near-infrared data at late times are essential to investigate the astrophysical origin of interesting transient objects. 
\item Assuming a BNS origin, our study suggests that this system was a 1.52$^{+0.49}_{-0.38} M_{\odot}$ - 1.30$^{+0.24}_{-0.32} M_{\odot}$ binary, leading to a total ejecta mass of $M^{\rm{BNS}}_{\rm ej, Kasen}~=~0.016^{+0.013}_{-0.009} \rm{M}_{\odot}$.
Assuming a NSBH origin of GRB~211211A, our study suggests a 3.11$^{+5.53}_{-2.23} M_{\odot}$ - 1.40$^{+0.74}_{-0.81} M_{\odot}$ system with a total ejecta mass of $M^{\rm{NSBH}}_{\rm ej}~=~0.006^{+0.006}_{-0.004}~M_{\odot}$.

\end{enumerate}

\label{sec:conclusion}

\section{Data availability}

The data underlying this article are available in the Zenodo repository at \url{https://zenodo.org/record/8284434}, doi: 10.5281/zenodo.8284434.

\section{Acknowledgements}
This project has received financial support from the Centre National de la Recherche Scientifique (CNRS) through Mission pour les initiatives transverses et interdisciplinaires (MITI) interdisciplinary programs. 
Sarah Antier thanks A. de Ugarte Postigo and J. Rastinejad for sharing data for this work. Sarah Antier also thanks Rahul Gupta, Jirong Mao, Robert Strausbaugh, Dong Xu, Jinzhong Liu, Daniele Malesani, Andrew Levan, and the Multicolor Imaging Telescopes for Survey and Monstrous Explosions (MITSuME) group for their useful comments on their observations. 
Mattia Bulla acknowledges support by the European Union’s Horizon 2020 Programme under the AHEAD2020 project (grant agreement n. 871158). 
Tim Dietrich acknowledges support from the Deutsche Forschungsgemeinschaft, DFG, project number DI 2553/7. Nina Kunert and Tim Dietrich acknowledge support from the Daimler and Benz Foundation for the project `NUMANJI'.
Co-funded by the European Union (European Research Council (ERC), SMArt, 101076369). Views and opinions expressed are those of the authors only and do not necessarily reflect those of the European Union or the ERC. Neither the European Union nor the granting authority can be held responsible for them.
Peter T. H. Pang is supported by the research program of the Netherlands Organization for Scientific Research (NWO).
Michael W. Coughlin and Brian Healy acknowledges support from the National Science Foundation with grant numbers PHY-2308862 and OAC-2117997.
The work of Ingo Tews was supported by the U.S. Department of Energy, Office of Science, Office of Nuclear Physics, under Contract No.~DE-AC52-06NA25396, by the Laboratory Directed Research and Development program of Los Alamos National Laboratory under Project No.\ 20220541ECR, and by the U.S. Department of Energy, Office of Science, Office of Advanced Scientific Computing Research, Scientific Discovery through Advanced Computing (SciDAC) NUCLEI program.
Shreya Anand acknowledges support from the National Science Foundation GROWTH PIRE grant No.\ 1545949. 
This work used Expanse at the San Diego Supercomputer Cluster through
allocation AST200029 -- "Towards a complete catalog of variable
sources to support efficient searches for compact binary mergers and
their products" from the Advanced Cyberinfrastructure Coordination
Ecosystem: Services \& Support (ACCESS) program, which is supported by
National Science Foundation grants \#2138259, \#2138286, \#2138307,
\#2137603, and \#2138296.

\bibliography{references}

\newcommand{\noop}[1]{}
\begin{thebibliography}{}
\expandafter\ifx\csname natexlab\endcsname\relax\def\natexlab#1{#1}\fi
\providecommand{\url}[1]{\href{#1}{#1}}

\bibitem[{{Abbott} {et~al.}(2017){Abbott}, {Abbott}, {Abbott}, {Acernese},
  {Ackley}, {Adams}, {Adams}, \& {others}}]{2017ApJ...848L..13A}
{Abbott}, B.~P., {Abbott}, R., {Abbott}, T.~D., {et~al.} 2017, \apjl, 848, L13

\bibitem[{Abbott {et~al.}(2017)Abbott, Abbott, Abbott, Acernese, Ackley, Adams,
  {et~al.}}]{LVCGW170817}
Abbott, B.~P., Abbott, R., Abbott, T.~D., {et~al.} 2017, Phys. Rev. Lett., 119,
  161101

\bibitem[{{Abbott} {et~al.}(2017){Abbott}, {Abbott}, {Abbott}, {Acernese},
  {Ackley}, {Adams}, {Adams}, {Addesso}, {Adhikari}, {Adya}, \&
  {Affeldt}}]{LSCMM2017ApJ}
{Abbott}, B.~P., {Abbott}, R., {Abbott}, T.~D., {et~al.} 2017, \apjl, 848, L12

\bibitem[{Abbott {et~al.}(2017)}]{LIGOScientific:2017adf}
Abbott, B.~P., {et~al.} 2017, Nature, 551, 85

\bibitem[{Abbott {et~al.}(2019)}]{LIGOScientific:2018mvr}
---. 2019, Phys. Rev. X, 9, 031040

\bibitem[{Abbott {et~al.}(2021)}]{LIGOScientific:2021djp}
Abbott, R., {et~al.} 2021, arXiv:2111.03606

\bibitem[{{Abbott, R. and others}(2021)}]{LIGOScientific:2020ibl}
{Abbott, R. and others}. 2021, Phys. Rev. X, 11, 021053

\bibitem[{{Ahumada} {et~al.}(2021){Ahumada}, {Singer}, {Anand}, {Coughlin},
  {Kasliwal}, {Ryan}, {Andreoni}, {Cenko}, {Fremling}, {Kumar}, {Pang},
  {Burns}, {Cunningham}, {Dichiara}, {Dietrich}, {Svinkin}, {Almualla},
  {Castro-Tirado}, {De}, {Dunwoody}, {Gatkine}, {Hammerstein}, {Iyyani},
  {Mangan}, {Perley}, {Purkayastha}, {Bellm}, {Bhalerao}, {Bolin}, {Bulla},
  {Cannella}, {Chandra}, {Duev}, {Frederiks}, {Gal-Yam}, {Graham}, {Ho},
  {Hurley}, {Karambelkar}, {Kool}, {Kulkarni}, {Mahabal}, {Masci}, {McBreen},
  {Pandey}, {Reusch}, {Ridnaia}, {Rosnet}, {Rusholme}, {Carracedo}, {Smith},
  {Soumagnac}, {Stein}, {Troja}, {Tsvetkova}, {Walters}, \&
  {Valeev}}]{2021NatAs...5..917A}
{Ahumada}, T., {Singer}, L.~P., {Anand}, S., {et~al.} 2021, Nature Astronomy,
  5, 917

\bibitem[{Anand {et~al.}(2021)}]{Anand:2020eyg}
Anand, S., {et~al.} 2021, Nature Astron., 5, 46

\bibitem[{Ascenzi {et~al.}(2019)}]{Ascenzi:2018mbh}
Ascenzi, S., {et~al.} 2019, Mon. Not. Roy. Astron. Soc., 486, 672

\bibitem[{Barnes \& Metzger(2022)}]{Barnes_2022}
Barnes, J., \& Metzger, B.~D. 2022, The Astrophysical Journal Letters, 939, L29

\bibitem[{Barnes \& Metzger(2023)}]{Barnes:2023}
---. 2023, Astrophys. J., 947, 55

\bibitem[{Bauswein {et~al.}(2017)Bauswein, Just, Janka, \&
  Stergioulas}]{Bauswein:2017vtn}
Bauswein, A., Just, O., Janka, H.-T., \& Stergioulas, N. 2017, Astrophys. J.
  Lett., 850, L34

\bibitem[{Beloborodov(2003)}]{Beloborodov:2002af}
Beloborodov, A.~M. 2003, Astrophys. J., 588, 931

\bibitem[{Berger {et~al.}(2013)Berger, Fong, \& Chornock}]{Berger:2013wna}
Berger, E., Fong, W., \& Chornock, R. 2013, Astrophys. J. Lett., 774, L23

\bibitem[{Buchner {et~al.}(2014)Buchner, Georgakakis, Nandra, Hsu, Rangel,
  Brightman, Merloni, Salvato, Donley, \& Kocevski}]{Buchner:2014nha}
Buchner, J., Georgakakis, A., Nandra, K., {et~al.} 2014, Astron. Astrophys.,
  564, A125

\bibitem[{Bulla(2019)}]{Bulla:2019muo}
Bulla, M. 2019, Mon. Not. Roy. Astron. Soc., 489, 5037

\bibitem[{Bulla(2023)}]{Bulla:2022mwo}
---. 2023, Mon. Not. Roy. Astron. Soc., 520, 2558

\bibitem[{Burbidge {et~al.}(1957)Burbidge, Burbidge, Fowler, \&
  Hoyle}]{RevModPhys.29.547}
Burbidge, E.~M., Burbidge, G.~R., Fowler, W.~A., \& Hoyle, F. 1957, Rev. Mod.
  Phys., 29, 547

\bibitem[{Capano {et~al.}(2020)Capano, Tews, Brown, Margalit, De, Kumar, Brown,
  Krishnan, \& Reddy}]{Capano:2019eae}
Capano, C.~D., Tews, I., Brown, S.~M., {et~al.} 2020, Nature Astron., 4, 625

\bibitem[{Coughlin {et~al.}(2019)Coughlin, Dietrich, Margalit, \&
  Metzger}]{Coughlin:2018fis}
Coughlin, M.~W., Dietrich, T., Margalit, B., \& Metzger, B.~D. 2019, Mon. Not.
  Roy. Astron. Soc., 489, L91

\bibitem[{Coughlin {et~al.}(2018)Coughlin, Dietrich, Doctor, Kasen, Coughlin,
  Jerkstrand, Leloudas, McBrien, Metzger, O’Shaughnessy, \&
  Smartt}]{Coughlin:2018miv}
Coughlin, M.~W., Dietrich, T., Doctor, Z., {et~al.} 2018, Monthly Notices of
  the Royal Astronomical Society, 480, 3871

\bibitem[{{Dietrich} {et~al.}(2020){Dietrich}, {Coughlin}, {Pang}, {Bulla},
  {Heinzel}, {Issa}, {Tews}, \& {Antier}}]{Dietrich2020}
{Dietrich}, T., {Coughlin}, M.~W., {Pang}, P. T.~H., {et~al.} 2020, Science,
  370, 1450

\bibitem[{Domoto {et~al.}(2022)Domoto, Tanaka, Kato, Kawaguchi, Hotokezaka, \&
  Wanajo}]{Domoto:2022cqp}
Domoto, N., Tanaka, M., Kato, D., {et~al.} 2022, Astrophys. J., 939, 8

\bibitem[{Farr {et~al.}(2011)Farr, Sravan, Cantrell, Kreidberg, Bailyn, Mandel,
  \& Kalogera}]{Farr_2011}
Farr, W.~M., Sravan, N., Cantrell, A., {et~al.} 2011, The Astrophysical
  Journal, 741, 103

\bibitem[{Fong {et~al.}(2016)Fong, Margutti, Chornock, Berger, Shappee, Levan,
  Tanvir, Smith, Milne, Laskar, Fox, Lunnan, Blanchard, Hjorth, Wiersema,
  van~der Horst, \& Zaritsky}]{FoMa2016}
Fong, W., Margutti, R., Chornock, R., {et~al.} 2016, The Astrophysical Journal,
  833, 151

\bibitem[{Foucart {et~al.}(2018)Foucart, Hinderer, \& Nissanke}]{Foucart_2018}
Foucart, F., Hinderer, T., \& Nissanke, S. 2018, Physical Review D, 98

\bibitem[{Gao {et~al.}(2022)Gao, Lei, \& Zhu}]{2022arXiv220505031G}
Gao, H., Lei, W.-H., \& Zhu, Z.-P. 2022, Astrophys. J. Lett., 934, L12

\bibitem[{{Gehrels} {et~al.}(2008){Gehrels}, {Barthelmy}, {Burrows},
  {Cannizzo}, {Chincarini}, {Fenimore}, {Kouveliotou}, {O'Brien}, {Palmer},
  {Racusin}, {Roming}, {Sakamoto}, {Tueller}, {Wijers}, \&
  {Zhang}}]{2008ApJ...689.1161G}
{Gehrels}, N., {Barthelmy}, S.~D., {Burrows}, D.~N., {et~al.} 2008, \apj, 689,
  1161

\bibitem[{Gompertz {et~al.}(2023)}]{2022arXiv220505008G}
Gompertz, B.~P., {et~al.} 2023, Nature Astron., 7, 67

\bibitem[{Henkel {et~al.}(2023)Henkel, Foucart, Raaijmakers, \&
  Nissanke}]{Henkel:2022naw}
Henkel, A., Foucart, F., Raaijmakers, G., \& Nissanke, S. 2023, Phys. Rev. D,
  107, 063028

\bibitem[{Huth {et~al.}(2022)}]{Huth:2021bsp}
Huth, S., {et~al.} 2022, Nature, 606, 276

\bibitem[{Jeffreys(1961)}]{Jeffreys1961}
Jeffreys, H. 1961, Theory of Probability (Oxford University Press)

\bibitem[{Jin {et~al.}(2020)Jin, Covino, Liao, Li, D'Avanzo, Fan, \&
  Wei}]{Jin:2019uqr}
Jin, Z.-P., Covino, S., Liao, N.-H., {et~al.} 2020, Nature Astron., 4, 77

\bibitem[{{Jin} {et~al.}(2015){Jin}, {Li}, {Cano}, {Covino}, {Fan}, \&
  {Wei}}]{JinLi2015}
{Jin}, Z.-P., {Li}, X., {Cano}, Z., {et~al.} 2015, \apjl, 811, L22

\bibitem[{{Jin} {et~al.}(2018){Jin}, {Li}, {Wang}, {Wang}, {He}, {Yuan},
  {Zhang}, {Zou}, {Fan}, \& {Wei}}]{JinWang2018}
{Jin}, Z.-P., {Li}, X., {Wang}, H., {et~al.} 2018, \apj, 857, 128

\bibitem[{Kasen {et~al.}(2017)Kasen, Metzger, Barnes, Quataert, \&
  Ramirez-Ruiz}]{Kasen:2017sxr}
Kasen, D., Metzger, B., Barnes, J., Quataert, E., \& Ramirez-Ruiz, E. 2017,
  Nature, 551, 80

\bibitem[{Kasen {et~al.}(2006)Kasen, Thomas, \& Nugent}]{Kasen:2006ce}
Kasen, D., Thomas, R.~C., \& Nugent, P. 2006, Astrophys. J., 651, 366

\bibitem[{{Kasliwal} {et~al.}(2017){Kasliwal}, {Korobkin}, {Lau}, {Wollaeger},
  \& {Fryer}}]{KaKoLau2017}
{Kasliwal}, M.~M., {Korobkin}, O., {Lau}, R.~M., {Wollaeger}, R., \& {Fryer},
  C.~L. 2017, \apjl, 843, L34

\bibitem[{Kass \& Raftery(1995)}]{Kass1995}
Kass, R.~E., \& Raftery, A.~E. 1995, Journal of the American Statistical
  Association, 90, 773

\bibitem[{Kr\"uger \& Foucart(2020)}]{Kruger:2020gig}
Kr\"uger, C.~J., \& Foucart, F. 2020, Phys. Rev. D, 101, 103002

\bibitem[{{Lamb} {et~al.}(2019){Lamb}, {Tanvir}, {Levan}, {de Ugarte Postigo},
  {Kawaguchi}, {Corsi}, {Evans}, {Gompertz}, {Malesani}, {Page}, {Wiersema},
  {Rosswog}, {Shibata}, {Tanaka}, {van der Horst}, {Cano}, {Fynbo}, {Fruchter},
  {Greiner}, {Heintz}, {Higgins}, {Hjorth}, {Izzo}, {Jakobsson}, {Kann},
  {O'Brien}, {Perley}, {Pian}, {Pugliese}, {Starling}, {Th{\"o}ne}, {Watson},
  {Wijers}, \& {Xu}}]{Lamb:2019lao}
{Lamb}, G.~P., {Tanvir}, N.~R., {Levan}, A.~J., {et~al.} 2019, \apj, 883, 48

\bibitem[{{Levan} {et~al.}(2016){Levan}, {Crowther}, {de Grijs}, {Langer},
  {Xu}, \& {Yoon}}]{2016SSRv..202...33L}
{Levan}, A., {Crowther}, P., {de Grijs}, R., {et~al.} 2016, \ssr, 202, 33

\bibitem[{Levan {et~al.}(2005)}]{Levan:2004sn}
Levan, A., {et~al.} 2005, Astrophys. J., 624, 880

\bibitem[{{Mao} {et~al.}(2021){Mao}, {Xin}, \& {Bai}}]{2021GCN.31232....1M}
{Mao}, J., {Xin}, Y.~X., \& {Bai}, J.~M. 2021, GRB Coordinates Network, 31232,
  1

\bibitem[{Mei {et~al.}(2022)}]{Mei:2022ncd}
Mei, A., {et~al.} 2022, Nature, 612, 236

\bibitem[{Most {et~al.}(2018)Most, Weih, Rezzolla, \&
  Schaffner-Bielich}]{Most:2018hfd}
Most, E.~R., Weih, L.~R., Rezzolla, L., \& Schaffner-Bielich, J. 2018, Phys.
  Rev. Lett., 120, 261103

\bibitem[{Nicholl {et~al.}(2021)Nicholl, Margalit, Schmidt, Smith, Ridley, \&
  Nuttall}]{Nicholl:2021rcr}
Nicholl, M., Margalit, B., Schmidt, P., {et~al.} 2021, Mon. Not. Roy. Astron.
  Soc., 505, 3016

\bibitem[{Pang {et~al.}(2022)}]{Pang:2022rzc}
Pang, P. T.~H., {et~al.} 2022, arXiv:2205.08513

\bibitem[{{Pankov} {et~al.}(2021){Pankov}, {Pozanenko}, {Belkin}, {Inasaridze},
  {Datashvili}, {Ayvazian}, {Kapanadze}, \& {GRB IKI
  FuN}}]{2021GCN.31233....1P}
{Pankov}, N., {Pozanenko}, A., {Belkin}, S., {et~al.} 2021, GRB Coordinates
  Network, 31233, 1

\bibitem[{Qian \& Woosley(1996)}]{Qian:1996xt}
Qian, Y.~Z., \& Woosley, S.~E. 1996, Astrophys. J., 471, 331

\bibitem[{Radice {et~al.}(2018)Radice, Perego, Zappa, \&
  Bernuzzi}]{Radice:2017lry}
Radice, D., Perego, A., Zappa, F., \& Bernuzzi, S. 2018, Astrophys. J. Lett.,
  852, L29

\bibitem[{Rastinejad {et~al.}(2022)}]{Rastinejad:2022zbg}
Rastinejad, J.~C., {et~al.} 2022, Nature, 612, 223

\bibitem[{Rossi {et~al.}(2022)}]{Rossi:2021bau}
Rossi, A., {et~al.} 2022, Astrophys. J., 932, 1

\bibitem[{Roth \& Kasen(2015)}]{Roth:2014wda}
Roth, N., \& Kasen, D. 2015, Astrophys. J. Suppl., 217, 9

\bibitem[{Ruiz {et~al.}(2018)Ruiz, Shapiro, \& Tsokaros}]{Ruiz:2017due}
Ruiz, M., Shapiro, S.~L., \& Tsokaros, A. 2018, Phys. Rev. D, 97, 021501

\bibitem[{Ryan {et~al.}(2020)Ryan, van Eerten, Piro, \& Troja}]{Ryan:2019fhz}
Ryan, G., van Eerten, H., Piro, L., \& Troja, E. 2020, Astrophys. J., 896, 166

\bibitem[{{Schlafly} \& {Finkbeiner}(2011)}]{SchaflyFinkbeiner2011}
{Schlafly}, E.~F., \& {Finkbeiner}, D.~P. 2011, \apj, 737, 103

\bibitem[{Stanek {et~al.}(2003)Stanek, Matheson, Garnavich, Martini, Berlind,
  Caldwell, Challis, Brown, Schild, Krisciunas, Calkins, Lee, Hathi, Jansen,
  Windhorst, Echevarria, Eisenstein, Pindor, Olszewski, Harding, Holland, \&
  Bersier}]{Stanek_2003}
Stanek, K.~Z., Matheson, T., Garnavich, P.~M., {et~al.} 2003, The Astrophysical
  Journal, 591, L17

\bibitem[{Suvorov {et~al.}(2022)Suvorov, Kuan, \&
  Kokkotas}]{2022arXiv220511112S}
Suvorov, A.~G., Kuan, H.-J., \& Kokkotas, K.~D. 2022, Astron. Astrophys., 664,
  A177

\bibitem[{Tanvir {et~al.}(2013)Tanvir, Levan, Fruchter, Hjorth, Wiersema,
  Tunnicliffe, \& de~Ugarte~Postigo}]{Tanvir:2013pia}
Tanvir, N.~R., Levan, A.~J., Fruchter, A.~S., {et~al.} 2013, Nature, 500, 547

\bibitem[{Troja {et~al.}(2018)}]{TrRy2018}
Troja, E., {et~al.} 2018, Nature Commun., 9, 4089

\bibitem[{Troja {et~al.}(2019)}]{Troja:2019ccb}
---. 2019, Mon. Not. Roy. Astron. Soc., 489, 2104, [Erratum:
  Mon.Not.Roy.Astron.Soc. 490, 4367 (2019)]

\bibitem[{Troja {et~al.}(2022)}]{Troja:2022yya}
---. 2022, Nature, 612, 228

\bibitem[{van Eerten {et~al.}(2010)van Eerten, Zhang, \&
  MacFadyen}]{vanEerten:2010zh}
van Eerten, H., Zhang, W., \& MacFadyen, A. 2010, Astrophys. J., 722, 235

\bibitem[{Watson {et~al.}(2019)}]{WaHa19}
Watson, D., {et~al.} 2019, Nature, 574, 497

\bibitem[{{Xiao} {et~al.}(2022){Xiao}, {Zhang}, {Zhu}, {Xiong}, {Gao}, {Xu},
  {Zhang}, {Peng}, {Li}, {Zhang}, {Lu}, {Lin}, {Liu}, {Zhang}, {Ge}, {Tuo},
  {Xue}, {Fu}, {Liu}, {Li}, {Wang}, {Zheng}, {Wang}, {Jiang}, {Li}, {Liu},
  {Cao}, {Cai}, {Yi}, {Zhao}, {Xie}, {Li}, {Luo}, {Liao}, {Song}, {Zhang},
  {Qu}, {Liu}, {Li}, {Xu}, \& {Li}}]{2022arXiv220502186X}
{Xiao}, S., {Zhang}, Y.-Q., {Zhu}, Z.-P., {et~al.} 2022, arXiv e-prints,
  arXiv:2205.02186

\bibitem[{{Yang} {et~al.}(2015){Yang}, {Jin}, {Li}, {Covino}, {Zheng},
  {Hotokezaka}, {Fan}, {Piran}, \& {Wei}}]{YaJi2015}
{Yang}, B., {Jin}, Z.-P., {Li}, X., {et~al.} 2015, Nature Communications, 6,
  7323

\bibitem[{Yang {et~al.}(2022)Yang, Ai, Zhang, Zhang, Liu, Wang, Yang, Yin, Li,
  \& L\"u}]{Yang:2022qmy}
Yang, J., Ai, S., Zhang, B.-B., {et~al.} 2022, Nature, 612, 232

\bibitem[{Ye \& Fishbach(2022)}]{Ye:2022qoe}
Ye, C., \& Fishbach, M. 2022, Astrophys. J., 937, 73

\bibitem[{Zhang {et~al.}(2021)}]{Zhang:2021agu}
Zhang, B.~B., {et~al.} 2021, Nature Astron., 5, 911

\bibitem[{{Zhang} {et~al.}(2022){Zhang}, {Huang}, {Zheng}, {Liu}, \&
  {Wang}}]{Zhang:2022fzj}
{Zhang}, H.-M., {Huang}, Y.-Y., {Zheng}, J.-H., {Liu}, R.-Y., \& {Wang}, X.-Y.
  2022, \apjl, 933, L22

\bibitem[{Özel {et~al.}(2010)Özel, Psaltis, Narayan, \&
  McClintock}]{ozel_2010}
Özel, F., Psaltis, D., Narayan, R., \& McClintock, J.~E. 2010, The
  Astrophysical Journal, 725, 1918

\end{thebibliography}

\clearpage

\appendix

\section{Inference settings}\label{App:Inf_results}

All parameter estimation runs were performed using the nuclear physics and multimessenger astronomy framework \texttt{NMMA}~\citep{Pang:2022rzc}. In this framework, joint Bayesian inferences of electromagnetic signals are carried out on the basis of the nested sampling algorithm implemented in \textsc{pymultinest} (\cite{Buchner:2014nha}). 
Each simulation used 2048 live points, and the prior settings for each of the employed models, as well as the median values and 90 per cent credible ranges, are provided in Table~\ref{tab:Prior_bounds_Dfix}.

\begin{table*}[h!]
  \centering
  \begin{tabular}{| m{2.8cm} | m{1.4cm}| m{2.2cm}| m{2.2cm}| m{2.2cm}| m{2.2cm}| m{2.2cm}|}
    \toprule
    \multirow{3}{*}{Parameter} & \multirow{3}{*}{Prior} & \multicolumn{5}{|c|}{Posterior} \\
    &     & BNS-GRB-M$^{\rm Bulla}_{\rm top}$ & BNS-GRB-M$^{\rm Kasen}_{\rm top}$ & NSBH-GRB-M$_{\rm top}$ &  SNCol-GRB-M$_{\rm top}$ & SN98bw-GRB-M$_{\rm top}$  \\ \hline
    \textbf{GRB-M}   &                         &       &        &       & &\\
    $\log_{\rm{10}} (E_{\rm K,iso})$ [erg] & [47, 55]                & $50.74^{+1.03}_{-0.86}$  & $50.63^{+0.90}_{-0.86}$ & $50.79^{+1.13}_{-0.94}$ & 50.72$^{+0.99}_{-0.89}$  & 50.40$^{+1.01}_{-0.70}$ \\
    $\theta_{\rm{Obs}}$ [rad]             & [0, $\frac{\pi}{4}$] & 0.02$^{+0.02}_{-0.02}$ & 0.02$^{+0.02}_{-0.02}$  & 0.02$^{+0.02}_{-0.02}$  & 0.02$^{+0.02}_{-0.02}$ & 0.04$^{+0.03}_{-0.02}$ \\ 
    $\theta_c$ [rad]                      & [0.01, $\frac{\pi}{10}$]& 0.03$^{+0.03}_{-0.02}$ & 0.03$^{+0.03}_{-0.02}$  & 0.03$^{+0.03}_{-0.02}$  & 0.03$^{+0.03}_{-0.02}$  & 0.06$^{+0.05}_{-0.03}$ \\ 
    $\log_{\rm{10}} (n)$ [$\rm{cm}^{-3}$] & [-6, 2]                 & -5.00$^{+1.46}_{-1.00}$ & -5.05$^{+1.29}_{-0.95}$ & -4.96$^{+1.58}_{-1.04}$  & -5.05$^{+1.30}_{-0.95}$& -4.77$^{+1.15}_{-1.23}$ \\
    $p$                                   & [2.01, 3]               & 2.42$^{+0.19}_{-0.18}$ & 2.47$^{+0.19}_{-0.18}$& 2.42$^{+0.19}_{-0.19}$& 2.45$^{+0.19}_{-0.17}$& 2.49$^{+0.18}_{-0.17}$ \\
    $\log_{\rm{10}} (\epsilon_e)$         & [-5, 0]                 & -0.10$^{+0.10}_{-0.22}$ & -0.09$^{+0.09}_{-0.17}$ & -0.10$^{+0.10}_{-0.24}$ & -0.09$^{+0.09}_{-0.19}$  & -0.08$^{+0.08}_{-0.18}$ \\ 
    $\log_{\rm{10}} (\epsilon_B)$         & [-10, 0]                & -0.66$^{+0.66}_{-1.43}$& -0.59$^{+0.59}_{-1.25}$ & -0.72$^{+0.72}_{-1.55}$  & -0.65$^{+0.65}_{-1.30}$ & -0.59$^{+0.59}_{-1.28}$ \\
    $D_L [\rm{Mpc}]$  & $\mathcal{N}(350, 2)$           & 350.13$^{+3.49}_{-3.63}$ & 350.04$^{+3.57}_{-3.69}$ & 350.32$^{+3.81}_{-3.56}$  & 350.28$^{+3.75}_{-3.47}$ & 350.00$^{+3.60}_{-3.77}$ \\
    \hline
    \textbf{BNS-KN-Bulla}  &           &    &   & & & \\
    
    $\log_{\rm{10}} (M^{\rm{ej}}_{\rm dyn})$  [$M_{\odot}$]  & [-3, -1] & -1.89$^{+0.66}_{-0.49}$ & & & &\\
    $\log_{\rm{10}} (M^{\rm{ej}}_{\rm wind})$ [$M_{\odot}$]  & [-3, -0.5] & -2.23$^{+0.48}_{-0.72}$ & & & &\\
    $\Phi$ [deg]                                             & [15, 75] & 69.78$^{+5.22}_{-16.11}$ & & & &\\
    \hline
    \textbf{BNS-KN-Kasen}                                           & & &                          &  &    &  \\
    $\log_{\rm{10}} (M_{\rm{ej}})$            [$M_{\odot}$]  & [-2.5, -1]& & -1.79$^{+0.35}_{-0.32}$    & &  &\\
    $\log_{\rm{10}} (v_{\rm{ej}})$            [$c$]          & [-1.8, -1]& & -0.92$^{+0.46}_{-0.45}$  & & &\\
    $\log_{\rm{10}} (X_{\rm{lan}})$                          & [-4.5, -1]& & -1.69$^{+0.69}_{-0.82}$ &    & &\\\hline
    \textbf{NSBH-KN-Bulla}                                           &           & & &  &     & \\
    $\log_{\rm{10}} (M^{\rm{ej}}_{\rm dyn})$  [$M_{\odot}$]  & [-3, -1]  & & & -2.56$^{+0.49}_{-0.44}$ & &  \\
    $\log_{\rm{10}} (M^{\rm{ej}}_{\rm wind})$ [$M_{\odot}$]  & [-3, -0.5]& & & -2.56$^{+0.50}_{-0.40}$ &  & \\ \hline
    \textbf{SNCol} & & &  &   &   & \\
    $M_{\rm{ej}}$ [$M_{\odot}$]                              & [0, 0.5]  & & & & 0.02$^{+0.06}_{-0.02}$ &    \\
    $M_{\rm{Ni}}$ [$M_{\odot}$]                              & [0, 0.03]   & & & & 0.00$^{+0.00}_{-0.00}$ & \\ 
    $v_{\rm{ej}}$ [$c$]                                      & 
[0, 0.21] & & & & 0.20$^{+0.01}_{-0.02}$  & \\
    $M_{\rm{rp}}$ [$M_{\odot}$]                              & [0, 0.05] & & & & 0.01$^{+0.02}_{-0.01}$ & \\
    $\Psi_{\rm{mix}}$                                        & [0, 0.9]    & & & & 0.84$^{+0.06}_{-0.13}$& \\\hline
    \textbf{SN98bw}                                           & & &                        &  &     &    \\
    $S_{\rm{max}}$                                          & [0, 60]& &        &   &   & 34.41$^{+24.54}_{-24.81}$ \\\hline
    
  \hline    
  \end{tabular}
  \caption{
  Model parameters and prior bounds employed in our Bayesian inferences. We report median posterior values at 90 per cent credibility from simulations that were run with Top-hat jet structure and with a narrow Gaussian luminosity distance prior $\mathcal{N}(\mu, \sigma)$, with mean $\mu = 350$ Mpc and standard deviation $\sigma = 2$ Mpc. We employ a conditional prior on the inclination angle depending on the jet core opening angle, $p(\theta_{\rm{Obs}} | \theta_{\rm{c}})$, using a truncated Gaussian distribution, $\mathcal{N}_T(\mu, \sigma)$, where $\mu = 0$ and $\sigma = \theta_{\rm{c}}$ .
  }
  \label{tab:Prior_bounds_Dfix}
\end{table*}

\newpage
\section{Inference results}\label{App:Inf_corners}

\begin{figure}[h!]
    \centering
    \includegraphics[width=0.95\textwidth]{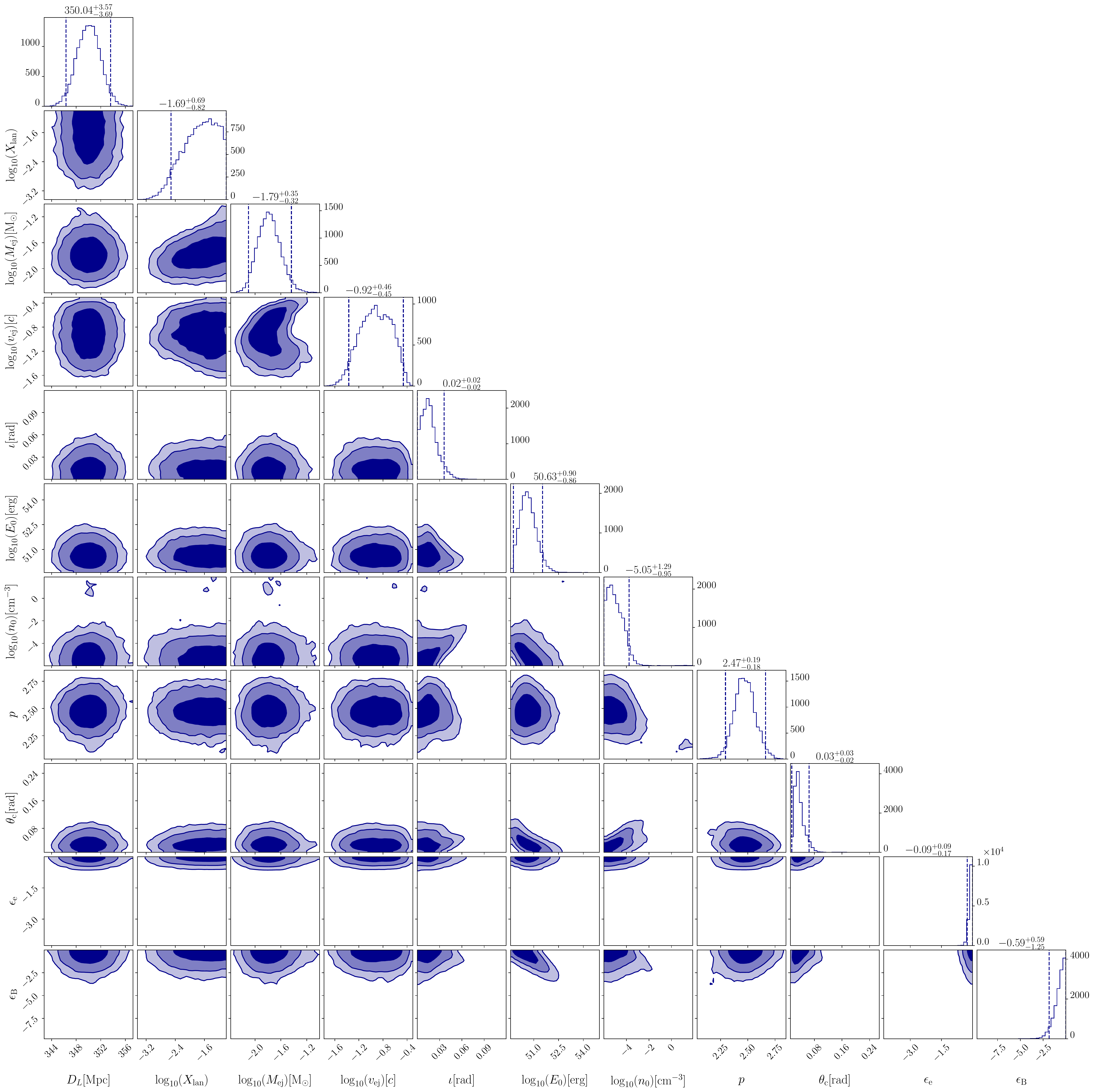}
    \caption{Corner plot for BNS-GRB-M$^{\rm{Kasen}}_{\rm{top}}$ with a narrow Gaussian luminosity distance prior centered around 350~Mpc, in which we show the inferred parameters at 68 per cent, 95 per cent, and 99 per cent confidence (shadings from light to dark). 
    For the 1D posterior probability distributions,
    we report the median values and show the 90 per cent confidence intervals as dashed lines.}
    \label{fig:Corner_BNSGRBKasenTophat}
\end{figure}

In the following, we present the posterior distribution for our reference model GRB$^{\rm{Kasen}}_{\rm{top}}$ employing a narrow distance prior centered around $350~\rm Mpc$. 
Figure~\ref{fig:Corner_BNSGRBKasenTophat} summarizes our results. As discussed in the main text, we obtain an average velocity of $10^{-0.92^{+0.46}_{-0.45}}\rm c$ and a total ejecta mass of $10^{-1.79^{+0.35}_{-0.32}}$~M$_\odot$, whereas the latter is slightly smaller as compared to previous findings in the literature. Interestingly, our analysis prefers a higher lanthanide fraction compared to the one inferred for AT2017gfo using the same kilonova models~\citep{Coughlin:2018miv}; that is, we predict a slightly redder kilonova (similar to~\cite{Rastinejad:2022zbg}, who predict a larger mass of the red component, but opposite to e.g.~\cite{Mei:2022ncd}).

\end{document}